\documentclass[lettersize,journal]{IEEEtran}

\usepackage{amsmath,amsfonts,amssymb}
\usepackage{orcidlink}

\usepackage{setspace}
\usepackage{notoccite}
\usepackage{algorithm}
\usepackage{algorithmic}

\usepackage{array}
\usepackage{titlesec}
\usepackage{textcomp}
\usepackage{stfloats}
\usepackage{url}
\usepackage{verbatim}
\usepackage{graphicx}

\usepackage{caption}
\usepackage{subcaption}

\usepackage{svg}
\usepackage{siunitx}
\usepackage{tabularx}
\usepackage{tikz}
\usepackage{pgfplots}
\usepackage[font=footnotesize]{caption}

\usepackage{balance}

\begin{document}
\title{Study of Multiuser Scheduling Based on User Satisfaction for MU-MIMO Systems} 
\author{Ahmed M. Nor \orcidlink{0000-0002-6943-892X},  Lukas T. N. Landau \orcidlink{0000-0003-4777-5885},   E. Veronica Belmega \orcidlink{0000-0003-4336-4704}  and Rodrigo C. de Lamare \orcidlink{0000-0003-2322-6451} \vspace{-0.1em}

%\IEEEauthorblockA{\textit{$^1$Department of Electrical Engineering, Pontifical Catholic University of Rio de Janeiro}, \text{Rio de Janeiro, Brazil} \\
%$^2$\textit{ Universit\'e Gustave Eiffel, CNRS, LIGM}, \text{Marne-la-Vallée, France}\\
%$^3$\textit{ETIS UMR 8051, CY Cergy Paris %Universit\'e, ENSEA, CNRS}, \text{Cergy, France}\\
%$^4$\textit{University of York}, \text{United Kingdom} \\
%ahmed.nor@esp.puc-rio.br; landau@puc-rio.br; veronica.belmega@esiee.fr; delamare@puc-rio.br}

\thanks{A. M. Nor, L. Landau and R. de Lamare are with the Department of Electrical Engineering, Pontifical Catholic University of Rio de Janeiro, Rio de Janeiro, Brazil. E. V. Belmega is with Universit\'e Gustave Eiffel, CNRS, LIGM, Marne-la-Vallée, France and ETIS UMR 8051, CY Cergy Paris Universit\'e, ENSEA, CNRS, Cergy, France. R. de Lamare is also with the University of York, United Kingdom. The emails are ahmed.nor@esp.puc-rio.br; landau@puc-rio.br; veronica.belmega@esiee.fr; delamare@puc-rio.br. This work is supported by FAPESP, MCTIC, and CGI in Brazil, under project no. 2023/00579-0, and in part, by the French National Research Agency (ANR-22-PEFT-0007) as part of France 2030 and the NF-FITNESS project.}}

\begin{comment}
    \markboth{Journal of \LaTeX\ Class Files,~Vol.~18, No.~9, September~2020}%
{How to Use the IEEEtran \LaTeX \ Templates}
\end{comment}

\maketitle

\linespread{1.03}
\begin{abstract}
%Multiuser - V: not fan of this repetition
Scheduling in multiuser multiple input multiple output (MU-MIMO) systems is essential for efficient resource allocation and overall performance enhancement. In this work, a multiuser scheduling problem is formulated to maximize the product of user equipments' (UEs) aggregate satisfactions, which maintains user fairness. %, meanwhile eliminating satisfied UEs from the scheduling to preserve resources for un-satisfied ones. 
Solving such a combinatorial problem using exhaustive search (EX), which requires evaluating all possible multiuser groups within a massive number of resource blocks (RBs), is prohibitive. Instead, we propose an efficient users' satisfaction based scheduling approach (US-SA). In our US-SA, a low dimension sub-grouping matrix is constructed {at each frame}, which is used to schedule the best multiuser group in each time slot; satisfied users are eliminated from the scheduling process. Our US-SA performs close to the optimal EX method in terms of satisfaction, transmitted data amount, spectral efficiency, latency, and fairness with lower computational cost. Moreover, our experiments demonstrate that the proposed scheme outperforms competing techniques.
\end{abstract}


\bstctlcite{IEEEexample:BSTcontrol}

\section{Introduction}
\label{Introduction}

The need for massive data rates, ultra-low latency, massive connectivity and reliability is key for 5G and beyond networks which will feature high traffic applications \cite{Lei2021}. Thus, adapting Multi-User Multiple-Input Multiple-Output (MU-MIMO) systems to simultaneously serve multiple user equipments (UEs) is needed to cope with these demands \cite{mmimo,wence}. In dense MU-MIMO networks, multiuser scheduling is vital to efficiently take advantage of the limited available system resources and improve overall performance. Scheduling approaches have been proposed and classified in wireless communications into channel-independent schemes, e.g., round robin (RR), which have low performance in dynamic channels conditions and heterogeneous traffic, and channel-dependent approaches, e.g., proportional fairness (PrF) \cite{PF_WCL}, %\cite{Nor2022}. 
Most prior works focused on the single user {per resource block} special case. In the multiuser case, power allocation is often beneficial to enhance performance \cite{tds,jrpa} and scheduling \cite{cesg,rra} becomes a difficult combinatorial problem, where users' grouping is required, i.e., selecting which users will be scheduled in the target resource block (RB). 

For multiuser proportional fairness (MU-PrF) in MU-MIMO systems, the authors in \cite{SPF} studied the scheduling on 3GPP LTE downlink. To reduce scheduling complexity and achieve a performance close to optimal MU-PrF, the authors in \cite{MU-MIMO_Sch_2} developed mCore+ approach, while \cite{MU-MIMO_Sch_3} suggested a deep reinforcement learning (DRL) and k-nearest neighbors based approach. In \cite{MU-MIMO_Sch_1}, an orthonormal subspace alignment scheduling algorithm is proposed for highly dense low-latency scenarios. These existing mechanisms \cite{SPF}, \cite{MU-MIMO_Sch_2}, \cite{MU-MIMO_Sch_3}, \cite{MU-MIMO_Sch_1} optimize the total sum rate while maintaining UEs fairness. The UE rate is considered as a metric for its satisfaction, however, this metric is neither effective nor fair. Indeed, some UEs may be scheduled due to their good channels, as this will maximize the sum rate, even though they do not need more RBs, which wastes valuable resources. Indeed, the actual received data amount and UEs latency, which are the relevant indicators for the satisfaction of the UEs, are ignored in previous works.

The approaches in \cite{sat_paper1} and \cite{sat_paper3} consider users' satisfaction in their scheduling, however, they neglect other metrics, including UEs fairness and latency, which are highly critical in 5G networks. The work in \cite{sat_paper1} focuses on a specific vehicle-to-everything single UE scenario, where safety traffic is scheduled first followed by non-safety traffic. In \cite{sat_paper3}, the authors define system satisfaction as the number of paired users over the number of users that achieve transmission requirements. This work assumes UEs grouping is already done and do not consider each UE demand individually. In the prior works, the users' satisfactions are not based on the UEs current required data amount, thus ignoring the allocated RBs.  %\textcolor{blue}{Our previous work in [] suggested optimizing the aggregate UEs satisfaction in a simplified MU-MIMO system, where it neglects excluding satisfied UEs from the scheduling, also it assumes multiuser groups construction has already done once.}

This work formulates a novel multiuser scheduling optimization problem (OP) that targets maximizing the users' satisfaction, while eliminating satisfied UEs from scheduling pool to preserve RBs for unsatisfied users. This OP is non-polynomial (NP) and cannot be solved using traditional schemes with massive UEs and RBs. Unlike existing works, the solution of this OP, by employing the exhaustive search (EX) method, depends on the current and already received demands alongside the channel state information (CSI) of the UEs. The EX, at each time slot, leads to prohibitive complexity and delay given the combinatorial number of possible multiuser groups. 
%solving it, within each time slot using an exhaustive search (EX), given the combinatorial large number of possible multiuser groups, leads to high computational complexity and delay. 

To overcome this issue, we propose a novel users' satisfaction based scheduling approach (US-SA) in which a sub-grouping matrix of unsatisfied users is constructed, then the system only searches through these multiuser groups to select the best one at each time slot. When a certain UE reaches its satisfaction, it is eliminated from the scheduling, and the sub-grouping matrix is reconstructed. Our US-SA depends on current and historical users' satisfaction to achieve high aggregate users' satisfaction and maintain fair RBs allocation. Hence, 
%it achieves the OP objective, 
our method is able to reach a high performance, in terms of spectral efficiency and fairness among UEs, close to the EX performance with extremely lower complexity, e.g., 16-times lower in 8 UEs case. Moreover, our experiments demonstrate that it outperforms other benchmarks and closely approaches EX in terms of average UEs satisfaction, transmitted data amount, and latency even with increasing UEs number.
\begin{comment}
    The paper is organized as follows: the system model and the proposed multiuser scheduling problem formulation are introduced in Section \ref{System Model} and Section \ref{Problem Formulation}, respectively. Section \ref{Proposed Scheme} discusses the proposed US-SA, while Section \ref{Analysis} analyzes the US-SA and other techniques. Numerical results are presented in Section \ref{Numerical Results} and Section \ref{Conclusion} concludes the paper.
\end{comment}

\section{System Model}

\label{System Model}
\begin{figure}
    \centering
    \includegraphics[scale=0.875]{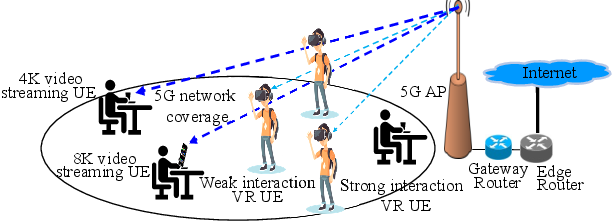}
    \setlength\abovecaptionskip{0\baselineskip}
    \caption{MU-MIMO network and its applications.}
    \label{system model_fig}
\end{figure}

A downlink multiuser scheduling problem in a MU-MIMO network \cite{mmimo,wence} %with enhanced mobile broadband (eMBB) services in 5G and beyond network 
is considered as in Fig \ref{system model_fig}, in which UEs experience applications with heterogeneous traffic, e.g., weak and strong entry-level virtual reality (VR), and 4K and 8K video streaming \cite{Lei2021}. A 60 GHz base station (BS) with $M$ antennas and 4 RF chains simultaneously serves up to $K_u$ UEs out of $K$ UEs associated to it, where $K_u \leq K$. Meanwhile, the UEs antennas are quasi-omnidirectional to reduce size and cost in practical systems. The BS performs all system operations and collects required information for UEs scheduling process, where perfect CSI knowledge is assumed. Also, if multi-cell scenario is considered, BSs can separately perform scheduling process. We also assume time-slotted scheduling, hence, within each time slot $t$, a group of up to $K_u$ users should be allocated.

Denote by $c_{k,j} \in \{0,1\}$ a variable indicating whether user $k \in K$ is a member of group $j$, where $c_{k,j}=1$ if $k$th user belongs to group $j$ and $c_{k,j}=0$ otherwise.
\begin{comment}
    \begin{equation}
c_{k,j} =
\begin{cases} 
1, & \text{if } k \text{th user exists in group } j, \\ 
0, & \text{otherwise.}
\end{cases}
\end{equation}
\end{comment}
Hence, the following constraint holds for $c_{k,j}$:
\begin{equation}
\sum_{k \in K} c_{k,j} \leq K_u, \forall j \in J,\tag{\textbf{C1}}
\label{C1}\end{equation}
%\begin{equation}
%\sum_{j \in J} c_{k,j} = 1, \forall k \in K,
%\tag{\textbf{C2}}
%\label{C2}\end{equation}
%\begin{equation}
%\textbf{C2: }\sum_{j \in J} c_{k,j} = 1, \quad \forall k \in K,
%\label{C2}\end{equation}
ensuring that each group contains {up to} $K_u$ UEs, and $J=\sum_{q=1}^{K_u}\binom{K}{q}$ is the number of multiuser group combinations.
%and each UE belongs to only one group

Let us denote $\gamma_j(t) \in \{0,1\}$ the variable indicating whether group $j \in J$ is scheduled in time slot $t$, where $\gamma_j(t)=1$ if $j$th group is scheduled in time slot $t$ and $\gamma_j(t)=0$ otherwise.
\begin{comment}
    \begin{equation}
\gamma_j(t) =
\begin{cases} 
1, & \text{if } j \text{th group is scheduled in time slot } t, \\ 
0, & \text{otherwise.}
\end{cases}
\end{equation}
\end{comment}
The following constraint holds for $\gamma_j(t)$:
\begin{equation}
\sum_{j \in J} \gamma_j(t) = 1, \forall t \in T,
\tag{\textbf{C2}}
\notag
\label{C2}\end{equation}
\begin{comment}
    \begin{equation}
\textbf{C4: }\sum_{t \in T} \gamma_j(t) \leq T, \quad \forall j \in J,
\label{C4}\end{equation}
\textbf{C4} indicates the limited RBs in the system.
\end{comment}
to ensure scheduling only one group in each slot $t$. $T=N_fN_s$ where $N_f$ is frames' number in each window size $t_w$, assuming a static channel and that each frame contains $N_s$ slots. 

Denote by $s_{k,j}(t)$ the ratio between the data amount in bits that $k$th UE can receive $d_{k,j}$, when it is a member of group $j$ and this group is scheduled in $t$th slot, and its required data $D_k^{\text{req}}(t)$ in $t$th slot. Thus, $s_{k,j}(t)$ can be expressed as
\begin{equation}
s_{k,j}(t) =
\begin{cases} 
{d_{k,j}}/{D_k^{\text{req}}(t)}, & D_k^{\text{req}}(t) > 0, \\ 
0, & \text{otherwise,}
\end{cases}
\label{UE_sat}\end{equation}
where $d_{k,j}$ can be computed based on Shannon's capacity as
\begin{equation}
    d_{k,j} =  t_{slot} \eta_{B} B \log_2(1+\text{SINR}_{k,j}),
\end{equation}
where $t_{slot}$ is one time slot duration. $\eta_{\text{B}}$ is bandwidth efficiency and $B$ is the channel bandwidth. $\text{SINR}_{k,j}$ is $k$th UE signal-to-noise plus interference ratio when it is allocated into group $j$, and can be expressed as
\begin{equation}
    \text{SINR}_{k,j} = \frac{c_{k,j}|\textbf{h}_{k}^H \textbf{w}_k |^2 P_{\max}}{\sum_{i=1, i\neq k}^{K} c_{i,j} |\textbf{h}_{k}^H \textbf{w}_i |^2 P_{\max} + \sigma_k^2 },
\end{equation}
%\textcolor{red}{[VB: Could we please explicitly include the transmit power explicitly in the SINR expression as above? ]}\\
where $\textbf{h}_{k}$ is the channel between the BS and $k$th UE, $\bf{w}_k = \tilde{\bf{w}}_k/ \|\tilde{\bf{w}}_k\|$ is normalized beamforming (BF) vector at the BS for $k$th user and $\bf{W} = [\tilde{\bf{w}}_1, \tilde{\textbf{w}}_2, ..., \tilde{\textbf{w}}_K]$. $P_{\max}$ is the maximum transmit power assuming equal power allocation, and $\sigma_k^2$ is the noise power. Zero-forcing (ZF) linear precoding is considered for BF to eliminate multi-user interference under the assumed channel knowledge and system dimension. Other precoding techniques can be considered \cite{gbd,wlbd,mbthp}.  ZF has a simple closed form $\textbf{W} = {\textbf{H}}({\textbf{H}}^H{\textbf{H}})^{-1}$ \cite{MU-MIMO_Sch_3}, where ${\textbf{H}} = [{\textbf{h}}_1, {\textbf{h}}_2, ..., {\textbf{h}}_K]$.
\begin{comment}
    \begin{equation}
    \textbf{W} = {\textbf{H}}({\textbf{H}}^H{\textbf{H}})^{-1}.
\end{equation}
\end{comment}
%\textcolor{red}{[VB: the above needs to be adjusted $\bf{w}_k = \tilde{\bf{w}}_k/ \|\tilde{\bf{w}}_k\|$, where the $\tilde{\bf{w}}_k$ are the columns of the matrix $\bf{W} = [\tilde{\bf{w}}_1, \tilde{\textbf{w}}_2, ..., \tilde{\textbf{w}}_K]$ which is computed as in equation (9).]}\\
Meanwhile, $D_k^{\text{req}}(t)$ can be defined as
\begin{equation}
D_k^{\text{req}}(t) =\max\{D_k^{\text{req}}(t-1)- \sum_{j \in J}d_{k,j}\gamma_j(t), 0\},
\label{data_req}
\end{equation} where $D_k^{\text{req}}(0) = R_k^{\text{req}} t_w$, and $R_k^{\text{req}}$ is the user $k$ required data rate. Hence, the $k$th UE satisfaction in time slot $t$ is
\begin{equation}
S_k (t) =
\begin{cases} 
c_{k,j}s_{k,j} (t) \gamma_{j}(t), & \text{if $j=j^*$}, \\ 
0, & \text{otherwise}.
\end{cases}
\end{equation}
where $j^*$ is the selected group that will be scheduled in time slot $t$. The $k$th UE aggregate satisfaction at the $t$th slot $S_k'(t)$ is the $k$th cumulative UE satisfaction: $S_k'(t) = \sum_{\tau=1}^{t} S_k (\tau)$. The $k$th UE becomes fully satisfied when $S_k'(t)=1$.

Let $\beta_k$ indicate whether UE is fully satisfied, then $\beta_k=1$, or not so $\beta_k=0$.
\begin{comment}
    \begin{equation}
\beta_k =
\begin{cases} 
0, & \text{if $k$th user is not satisfied}, \\ 
1, & \text{if $k$th user is satisfied}.
\end{cases}
\label{UE_sat_ind}\end{equation}
\end{comment}
Thus, the users' satisfaction indicator vector $\pmb{\beta}_K$ can be defined as $[\beta_1, \beta_2, ...,\beta_K]$.
$S_k'(t)$ can be written as
\begin{comment}
    \begin{multline}
    S_k'(t) = \sum_{\tau=1}^{t} S_k (\tau) = S_k (1)+S_k (2)+...+S_k (t-1)+S_k (t) \\
    = \sum_{\tau=1}^{t-1} S_k (\tau) + S_k (t) = S_k'(t-1) + S_k (t).
\label{agg_sat}
\end{multline}
\end{comment}
\begin{equation}
S_k'(t) =
\begin{cases} 
S_k'(t-1) + S_k (t), & \text{if }\beta_k = 0, \\ 
1, & \text{otherwise}.
\end{cases}
\label{agg_sat}
\end{equation}$S_k'(t)$ is modeled based on UEs quality of service, i.e., their required data amounts that depend on their various experienced service-type and indeed it is dynamically updated.
\begin{comment}
    \begin{figure}
    \centering
    \includegraphics[scale=0.55]{Content/Figures/fig_system_model.eps}
    \setlength\abovecaptionskip{0\baselineskip}
    \caption{MU-MIMO network and its applications.}
    \label{system model_fig}
\end{figure}
\end{comment}

\section{Proposed Problem Formulation}
\label{Problem Formulation}
The objective is to optimize the system performance by maximizing users' satisfaction and fairly distributing RBs with exclusion of satisfied users to preserve RBs for UEs in need. %, which indicates to the beneficial received data amount that the users gain when they are scheduled, with fair resources allocation consideration, under constraint that satisfied users will be eliminated from the scheduling process to preserve RBs for users in need. 
Thus, we propose maximizing the product of final aggregate users’ satisfaction $S_k'(T)$ as opposed to the sum of user satisfaction, which is not fair, aiming to fully satisfy all UEs. The optimal satisfaction happens when $\prod_{k=1}^{K} S_k'(T)=1$, where $S_k'(T) \leq 1$, $\forall k \in K$. The case of $S_k'(T) > 1$ will not occur in practice due to the non-existence of data to transmit to the $k$th UE if this UE achieves its satisfaction. The logarithm of $\prod_{k=1}^{K} S_k'(T)$ is strictly increasing. Hence, using the log identity, the resulting OP can be defined as \cite{MU-MIMO_Sch_2}
\begin{equation}
(P1) \quad \max_{\pmb\Gamma_J} \prod_{k=1}^{K} S_k'(T) \equiv \max_{\pmb\Gamma_J} \sum_{k=1}^{K} \log S_k'(T),
\label{P1_eq}   
\end{equation}
\[
\text{s.t.} \quad 0 < S_k'(T) \leq 1, \quad \forall k \in K,
\]
\[
c_{k,j} \in \{0,1\},(\textbf{C1}),\gamma_j(t) \in \{0,1\},(\textbf{C2}),
\]
where the scheduling matrix $\pmb\Gamma_J = [\gamma_j(t)]_{t,j}\in \mathbb{C}^{T \times J}$. The geometric mean is chosen for this OP due to its sensitivity to low UEs satisfactions, hence ensuring fairness among UEs. It is also less sensitive to outliers thus more reliable to achieve better performance in dynamic environments.
\begin{comment}
  Maximizing the product of satisfaction is equivalent to maximizing the sum of the logarithms of the users’ satisfaction, because the logarithm is a monotonically increasing.   
\end{comment}

In dense scenarios, optimally performing scheduling over all time RBs is highly complex and NP hard. Hence, it is more convenient to solve $(P1)$ by taking the scheduling decision at each time slot $t$. For this, we use the first-order Taylor series to approximate $\log S_k'(t)$ around $S_k' (t-1)$, where $S_k'(0)=\epsilon$ and $\epsilon \to 0^+$ is a small positive constant, as in \cite{Taylor}
\begin{equation}
\log S_k' (t) \approx \log S_k' (t-1) + \frac{S_k' (t) - S_k' (t-1)}{S_k' (t-1)} .
\label{taylor_exp}
\end{equation}
By substituting (\ref{agg_sat}) into (\ref{taylor_exp}), we can rewrite (\ref{taylor_exp}) as
\begin{equation}
\log S_k' (t) \approx \log S_k' (t-1) + S_k (t)/{S_k' (t-1)}.
\end{equation}
Then, by performing the summation over all users $K$, we get
\begin{equation}
\sum_{k=1}^{K} \log S_k' (t) \approx \sum_{k=1}^{K} \log S_k' (t-1) + \sum_{k=1}^{K} \frac{S_k (t)}{S_k' (t-1)}.
\label{taylor_sum}
\end{equation}
The term $\sum_{k=1}^{K} \log S_k' (t-1)$ is constant with respect to the scheduling decision. Thus, our scheduling problem at each slot $t$, while eliminating UEs which are satisfied, becomes:
\begin{equation}
\max_{\pmb\gamma_J(t)} \sum_{k=1}^{K} \frac{S_k (t)}{S_k' (t-1)}, \quad \forall t \in T,
\label{P1_@t}
\end{equation}
where $\pmb\gamma_J(t)\in \mathbb{C}^{1 \times J}$ is the scheduling vector at time slot $t$.

The new objective depends on UEs current satisfactions and their historical aggregate satisfactions $S_k'(t-1)$. This balances the optimization and maintains fairness, where RBs are allocated to the better conditions' UEs in the short term and the already served ones over the long term to improve their satisfactions. Hence, at each time slot $t$, optimal multiuser group $j^*$, that can be scheduled in this slot, is selected as 
\begin{comment}
    If only $S_k (t)$ is considered, users, who have high $S_k (t)$ and might been allocated many resources in the past, may continue be allocated in the current time slot and receiving more data, though they are near to their full satisfaction. On the other side, users with high needs maybe ignored and take a longer time to be more satisfied. Noting that this is a sub-optimal scheduling solution till certain time slot $\tau$, and some UEs might have temporarily better conditions, e.g., better satisfaction, allowing them to be allocated more resources in the short term. 
\end{comment}
\begin{equation}
j^*=\arg\max_{j \in J} \sum_{k \in K} \frac{s_{k,j} (t) c_{k,j}}{S_k' (t-1)} , \forall t \in T.
\label{P1_4algorithms}
\end{equation}
Then, the scheduling decision variables will be assigned as $\gamma_{j^*}(t)=1$ and $\gamma_{j \neq j^*}(t)=0$. The priority function of $j$th group $PF_j=\sum_{k \in K} \frac{s_{k,j} (t) c_{k,j}}{S_k' (t-1)}$. The intermediate variables $s_{k,j} (t)$ and $S_k'(t-1)$ are calculated using \eqref{UE_sat} and \eqref{agg_sat}. 
\begin{comment}
    \textcolor{red}{and we will refer to $\frac{s_{k,j} (t)}{S_k' (t-1)}c_{k,j}\gamma_j(t)$ as $PF_k$ which is the priority function of the $k$th user}.
\end{comment}
%In \eqref{P1_4algorithms}, $\sum_{k \in K} \frac{s_{k,j} (t)}{S_k' (t-1)} c_{k,j}$ is the $j$th group priority function $PF_j(t)$. 
Passing through all slots, $\pmb\Gamma_J$ will be determined. $c_{k,j}$ is an element in the grouping matrix $\textbf{C} \in \mathbb{C}^{J \times K}$ that contains all possible multiuser group combinations. To optimally solve \eqref{P1_4algorithms}, an EX over all $J$ groups, that exist in $\textbf{C}$, is performed. Employing EX in this combinatorial problem, which depends on $K$, $K_u$ and $J$, causes long delay and high complexity, alternatively, a cost-effective algorithm will be proposed in the next section. 

%This combinatorial problem depends on $K$, $K_u$ and $J = \sum_{q=1}^{K_u}\binom{K}{q} $. Hence, in dense networks, employing EX will cause long delay and high complexity. 

\section{Proposed US-SA}
\label{Proposed Scheme}
%In this section, we introduce US-SA a novel scheme for multiuser scheduling in MU-MIMO systems. % To relax EX based approach complexity, the multiuser groups' number, i.e., combinations, that the system looks into them to perform scheduling at each $t$ slot shall be reduced. Hence, 
%US-SA randomly constructs a sub-grouping matrix out of all possible combinations, then searches only through those groups to select the best one. In addition, the US-SA simultaneously eliminates satisfied UEs to reduce searching cost and rapidly maximize the objective. 
In this section, we introduce US‑SA, a novel scheme for multiuser scheduling in MU‑MIMO systems \cite{raus}. Instead of evaluating all possible UEs groupings, US‑SA randomly constructs a subset of candidate groups, represented as a sub‑grouping matrix, and searches in this reduced set to identify the best one. Additionally, US‑SA progressively eliminates UEs that are fully satisfied, thereby reducing the search space and accelerating the maximization of the objective function.

First, US-SA initializes the unsatisfied UEs number $K^*$.
To distribute $K$ UEs into multiuser groups, US-SA sets $L = K / K_u$, where each group contains at most $K_u$ users and no user is repeated across groups. Then, it initializes $\pmb\Gamma_L=\pmb{0}$, $\textbf{s}_K'=[S_1'(0),S_2'(0),...,S_K'(0)]=\pmb{\epsilon}$, and $\pmb{\beta}_K=\textbf{0}$. Also, it defines the set of unsatisfied users $\mathcal{U}_c = \{1,2,..., K\}$ and calculates the users' current required data amount vector $\textbf{d}_{K}=[D_1^{\text{req}}(0),D_2^{\text{req}}(0),...,D_K^{\text{req}}(0)]$ using \eqref{data_req}. 

At each new frame, the sub-grouping matrix $\mathbf{C}' \in \mathbb{C}^{L \times K}$ is randomly constructed as 
\begin{equation}
c_{k,\ell}'=
\begin{cases} 
1, & \text{if }k \in \text{Random Subset}\big(\mathcal{U}_\ell, u_\ell\big), \\ 
0, & \text{otherwise}, 
\end{cases}
\end{equation}
where $c_{k,\ell}'$ indicates whether or not user $k$ is a member of group $\ell$, where $\mathcal{U}_1 = \mathcal{U}_c$, then $\mathcal{U}_\ell=\mathcal{U}_{\ell-1} \setminus \text{RandomSubset}(U_{\ell-1}, u_{\ell-1}) \text{ for } \ell=2,.., L$. 
RandomSubset$\big(\mathcal{U}_\ell, u_\ell\big)$ randomly selects a $u_\ell$ users from $\mathcal{U}_\ell$ without replacement, where $u_\ell=\text{min}(K_u,|\mathcal{U}_\ell|)$. The following constraints exist for $c_{k,\ell}'$:
%$u_\ell$ is the number of users in multiuser group if the number of unsatisfied users is less than $K_u$.
{\fontsize{9.8}{11}\selectfont\begin{equation}
\sum_{k \in K} c_{k,\ell}' \leq K_u, \forall \ell \in L \ \ (\textbf{C3}),\quad\sum_{\ell \in L} c_{k,\ell}' = 1,  \forall k \in K \ \ (\textbf{C4})
\notag
\label{C4}\end{equation}}where (\textbf{C3}) is similar to (\textbf{C1}), and (\textbf{C4}) ensures each UE belongs to only one group. This frequent random grouping strategy ensures exploring the full multiuser grouping space instead of relying on a single configuration, which may not yield long term benefits. Moreover, randomly generated UEs grouping reduces system complexity and implementation, in future extension, alternative methods, e.g., correlation based UEs grouping, will be considered. Indeed, $\mathbf{C}'$ highly reduces the system complexity as compared to EX since $L \ll J$.

%\begin{equation}
%\textbf{C6: }\sum_{\ell \in L} c_{k,\ell}' = 1, \quad \forall k \in K,
%\label{C6}\end{equation}
Over time, $k$th UE priority function $PF_{k,\ell}$ is
\begin{equation}
PF_{k,\ell}(t)=
\begin{cases} 
\frac{s_{k,\ell} (t)}{S_k' (t-1)} c'_{k,\ell}, & \text{if } c'_{k,\ell} = 1, \beta_k=0, \\
0, & \text{otherwise.}
\end{cases}\label{PF_k}
\end{equation}
Then, for each $\ell$th group, US-SA calculates its priority function $PF_\ell(t)=\sum_{k \in K} PF_{k,\ell}(t) $, %Thus, the system gets the $PF_\ell(t)$ for all $L$ groups. 
then selecting the multiuser group $\ell^*$ and schedule this group in the $t$th time slot as
\begin{equation}
\ell^*=\arg\max_{\ell \in L} PF_\ell, \quad \forall t \in T,
\label{l_star}
\end{equation}
thus, $\gamma_{\ell^*}(t)=1$ and $\gamma_{\ell\neq \ell^*}(t)=0$. Also, the system updates the $\textbf{s}_K'$ and $\textbf{d}_{K}$ using \eqref{agg_sat} and \eqref{data_req}, respectively. Then, US-SA checks for UEs satisfactions, where fully satisfied UEs will be excluded from $\mathcal{U}_c$, and other parameters will be updated. $K^*=K^*-1$, $L=\text{ceil}(K^*/K_u)$, and $\mathbf{C}'$ will be reconstructed accordingly. US-SA will be terminated if all UEs are satisfied to save energy and relax the system. The US-SA is summarized in Algorithm \ref{proposed algorithm} and its simplified version (Sim. US-SA) can be used by discarding steps 13-20, which are designed for eliminating satisfied UEs. However, this will degrade performance, as shown in the results section.

\begin{algorithm}[!t]
\caption{Proposed US-SA.}\label{proposed algorithm}
\small
\textbf{Input: } $K$, $R_k^{\text{req}}$ for all $K$ users.\\
\textbf{Output: }Determine $\pmb\Gamma_L$.
\begin{algorithmic}[1]
\STATE \textbf{Initialize} $K^*=K$, $L=K/K_u$, $\pmb\Gamma_L=\pmb{0}$, $\textbf{s}_K'=\pmb{\epsilon}$, $\pmb{\beta}_K=\pmb{0}$,  and $\mathcal{U}_c=\{1,2,..., K\}$.
\STATE \textbf{Calculate} $\textbf{d}_{K}$ using \eqref{data_req}.
\FOR{$n_f=1...N_{f}$}{
\STATE \textbf{Construct} randomly sub-grouping matrix $\mathbf{C}'$ from $\mathcal{U}_c$.
\FOR{$n_s=1...N_s$}{
\STATE $t=(n_f-1)N_s+n_s$.
\FOR{$\ell=1...L$}{
%\FOR{$k=1...K$}{
%\STATE \textbf{Calculate} $PF_{k,\ell}(t)$ using \eqref{PF_k}.
%}\ENDFOR
\STATE \textbf{for} $k=1...K$ \textbf{Compute} $PF_{k,\ell}(t)$ using \eqref{PF_k} \textbf{end for}
\STATE \textbf{Calculate} $PF_\ell(t)=\sum_{k \in K} PF_{k,\ell}(t). $
}\ENDFOR
\STATE \textbf{Assign} current time slot to $\ell^*$ based on \eqref{l_star}.
\STATE \textbf{Update} $\textbf{s}_K'$ and $\textbf{d}_{K}$ using \eqref{agg_sat} and \eqref{data_req}.
\FOR{$k=1...K$}
\IF{$S_k'(t)= 1$ and $\beta_k=0$}
\STATE $\beta_k = 1$, $K^*=K^*-1$, and $L=\text{ceil}(K^*/K_u)$.
\STATE \textbf{if} $K^*=0$ \textbf{then \, Break; \, end if}
\STATE \textbf{Exclude} satisfied $k$th UE from $\mathcal{U}_c$.
\STATE \textbf{Construct} randomly $\mathbf{C}'$ from $\mathcal{U}_c$.
\ENDIF
\ENDFOR
}\ENDFOR
}\ENDFOR
\end{algorithmic}
\end{algorithm}

\section{Analysis}
\label{Analysis}
%In this section, we carry out a mathematical analysis of US-SA along with an evaluation of in its computation cost.
%\subsection{Mathematical Connections and Bounds}
A search in multiuser groups can be interpreted as a tree search \cite{TreeSearch_WCL}, where $\mathbf{C}$ represents a tree, each multiuser group is a path, each groups' combination is a sub-tree, each allocation of a UE into a group is a tree level, and the tree leaves indicate the allocation completion. 
The $\mathbf{C}'$ matrix in US-SA has $L$ groups out of $J$ groups that exist in \textbf{C}. Hence, in US-SA, at each frame, we can say one $J_s=J/L$ sub-tree is randomly chosen, consequently, this greatly reduces cost.

The computation complexity of US-SA is expressed in terms of $\mathcal{O}(.)$ and divided among its different stages. First, the initialization stage and $\textbf{d}_K$ calculation are $\mathcal{O}(LT+2K)$. Second, constructing the random sub-grouping at each frame costs $\mathcal{O}(LK)$, and calculating the $PF_{k,\ell}$ followed by $PF_{\ell}$ costs $\mathcal{O}(LK)$, assuming $PF_{k,l}$ costs $\mathcal{O}(1)$ per $k$th user. The assignment and the update steps cost $\mathcal{O}(L)$ and $\mathcal{O}(2K)$. Assuming all UEs will become fully satisfied, eliminating them and reconstructing $\mathbf{C}'$ will cost $\mathcal{O}(LK^2)$ as reconstructing $\mathbf{C}'$ can happen up to $K$ times, while un-satisfaction condition will occur the remaining $(T-K)$ times. Hence, the total complexity of US-SA is $\mathcal{O}(LT+LKN_f+T(LK+L+3K)+K^2(L-1))$. 
Table \ref{Complexity} summarizes all approaches' complexities. 

{RR has the lowest complexity but provides the worst performance. mCore+ and MU-PrF have higher complexities and consider different objective. EX achieves the best performance with the highest complexity. The proposed US-SA and Sim. US-SA outperform competing techniques and nearly approach EX with much lower complexities, hence US-SA can be implemented in practical systems.
US-SA has excellent performance due to: scheduling UEs based on their needs, constraining the search space to a limited number of groups, and groups' swapping at each frame ensures selection from different group configurations compensating for the randomness.}
\begin{table}[!t]
\setlength\tabcolsep{2pt}
\captionsetup{width=0.9\columnwidth}   
\setlength\belowcaptionskip{0\baselineskip}
\caption[]{Computational complexity.}
\label{Complexity}
\small
\begin{tabular}{|p{0.2\columnwidth}|p{0.79\columnwidth}|}
\hline
\textbf{Algorithm} & \textbf{Computational complexity}\\  
\hline
RR & $\mathcal{O}(T(K+1))$  \\
MU-PrF \cite{SPF} & $\mathcal{O}(JT+JK+T(JK+J+2K))$  \\
%Ref [conf] & $\mathcal{O}(LT+LK+T(LK+L+2K))$  \\
mCore+ \cite{MU-MIMO_Sch_2} & $\mathcal{O}(KT+TK\log(K)+T K^2_u M^2)$ \\
Sim. US-SA & $\mathcal{O}(LT+LKN_f+T(LK+L+2K))$  \\
US-SA & $\mathcal{O}(LT+LKN_f+T(LK+L+3K)+K^2(L-1))$  \\
EX & $\mathcal{O}(JT+JK+T(JK+J+3K)+K^2(J-1))$  \\
\hline
\end{tabular}
\end{table}

\section{Numerical Results}
\label{Numerical Results}
%This section evaluates the proposed US-SA against competing techniques, i.e., RR and MU-PrF \cite{SPF}, mCore+ \cite{MU-MIMO_Sch_2}, and EX based approach. The assessment is done through, the objective function value in \eqref{P1_eq}, the average users' satisfaction $S_K^{avg}$ and the probability of full users' satisfaction $P_{sat}^{full}$ that indicates the full satisfaction for all $K$ users:
%\begin{equation}
%S_K^{avg} = \frac{1}{K} \sum_{k=1}^K S'_k(T), \ \ 
%P_{sat}^{full}= \mathbb{P}\left(\prod_{k=1}^KS'_k(T)=1\right).
%\end{equation}
This section evaluates the proposed US-SA against competing techniques, i.e., RR, MU-PrF \cite{SPF}, mCore+ \cite{MU-MIMO_Sch_2}, deep reinforcement learning (DRL)-based \cite{MU-MIMO_Sch_3}, and EX based approaches. The assessment is done through the objective function value in \eqref{P1_eq}, normalized spectral efficiency (NSE) \cite{MU-MIMO_Sch_3} and the Jain's fairness index (JFI).
The normalization factor is the sum of the $K_{top}=4$ largest achievable rates out of the total $K$ UEs, where $k$th UE achievable rate is the transmitted data amount to this UE over the time required for this transmission if that UE is scheduled individually. JFI is expressed as
\begin{equation}
    JFI=\left(\sum_{k=1}^KS'_k(T)\right)^2 / \left(K\sum_{k=1}^K (S'_k(T))^2\right).
\end{equation}
We also present average users' satisfaction $S_K^{avg} = \frac{1}{K} \sum_{k=1}^K S'_k(T)$, the average transmitted data amount and the average latency, which indicates the required time before fully satisfying the UEs, i.e., all $K$ users obtain their required data.
\begin{comment}
   \begin{equation}
JFI=\left(\sum_{k=1}^KS'_k(T)\right)^2 / K\sum_{k=1}^K (S'_k(T))^2.
\end{equation} 
\end{comment}

The simulation parameters are given in Table \ref{sim_parameters}. We consider 8 to 40 UEs associated to the BS in our study to confirm the scalability of US-SA. All algorithms are run under full settings on a PC with specifications: Intel Core i7-10700 CPU, 32 GB RAM, 2.90 GHz, 4 Cores.
UEs are uniformly distributed at each realization. We averaged over $10^4$ trials to compensate for the randomness UEs grouping and channels, which is modeled based on 3GPP channel model for scene indoor mixed-office \cite{3GPP_TR38.901_channel}. In this scenario, UEs are static, hence, considering static channels are reasonable within the scheduling window. Also, UEs experience different applications, 4K and 8K UHD video streaming, that need $120$ and $300$ Mbps data rates; weak and strong interactions entry-level VR, that require $150$ and $200$ Mbps \cite{Lei2021}, assuming UEs experience these applications with $0.3$, $0.2$, $0.3$, and $0.2$ probabilities, respectively. Thus, e.g., the average required data amount for $16$ UEs case equals $16 (120\times0.3+150\times0.3+200\times0.2+300\times0.2) = 2.896$ Gbits.

\begin{figure*}[!t]
\centering
\begin{subfigure}[t]{0.3\textwidth}
    \centering
    \begin{tikzpicture}[scale = 0.55]
\begin{axis}[
    xlabel = {Time slot index},
    ylabel = {$P1$ objective function value},
    xtick distance=2000,
    ytick distance=0.1,
    xmin=2000,xmax=16001,
    ymin=0,ymax=1,
    scaled x ticks=base 10:-3,
    grid=major,
        yticklabel style={
        /pgf/number format/precision=2,
        /pgf/number format/fixed,
        /pgf/number format/fixed zerofill=true
    },
    label style={font=\large},
    legend entries={EX, Proposed US-SA, Proposed Sim. US-SA, MU-PrF, DRL based, mCore+, RR},
    legend style={
    draw=none,
    fill=white,
    fill opacity=0,
    text opacity=1,
    legend cell align=left,
    align=left,
    font=\normalsize, % 
    %draw=white!15!black,
    at={(0.62,0.613)}, % 
    anchor=south east,
    legend columns=1}
]
    \addplot [color=black, line width=2pt] 
    table [x=times,y=EX_res] {Content/Figures/OF_final.dat};
    \addplot [color=red, line width=2pt] 
    table [x=times,y=USSA_res] {Content/Figures/OF_final.dat};
    %\addplot [color=violet, line width=2pt] 
    %table [x=times,y=Proposed_conf_res] {Content/Figures/res_OF_let.dat};
    \addplot [color=cyan, line width=2pt] 
    table [x=times,y=Simplified_USSA_res] {Content/Figures/OF_final.dat};
    \addplot [color=green, line width=2pt] 
    table [x=times,y=MU_SPF_res] {Content/Figures/OF_final.dat};
    \addplot [color=orange, dotted, line width=2pt] 
    table [x=times,y=DRL] {Content/Figures/OF_final.dat};
    \addplot [color=violet, dashed, line width=2pt] 
    table [x=times,y=mCore_res] {Content/Figures/OF_final.dat};
    \addplot [color=blue, line width=2pt] 
    table [x=times,y=RR_res] {Content/Figures/OF_final.dat};
\end{axis}
\end{tikzpicture}
    \setlength\abovecaptionskip{0\baselineskip}
    \caption{}\label{fig_OF}
\end{subfigure}
\hfill
\begin{subfigure}[t]{0.3\textwidth}
    \centering
    \begin{tikzpicture}[scale = 0.55]
\begin{axis}[
    xlabel = {UEs number $K$},
    ylabel = {Normalized system spectral efficiency},
    xtick distance=4,
    ytick distance=0.02,
    xmin=14,xmax=18,
    ymin=0.2,ymax=0.40,
    grid=major,
    label style={font=\large},
    ybar=1pt,
    area legend,
    %precision=2,
    bar width=12pt,
    legend entries={EX, Proposed US-SA, Proposed Sim. US-SA, MU-PrF, DRL based, mCore+,  RR},
    legend style={
    fill=white,
    fill opacity=0.9,
    text opacity=1,
    legend cell align=left,
    align=left,
    font=\normalsize, % 
    draw=white!15!black,
    at={(0.61,0)}, % 
    anchor=south east,
    legend columns=1}
]
    \addplot [color=black, fill] coordinates {(16,0.385988)};
    \addplot [color=red, fill] coordinates {(16,0.379874)};
    \addplot [color=cyan, fill] coordinates {(16,0.365395)};
    \addplot [color=green, fill] coordinates {(16,0.35993)};
    \addplot [color=orange, fill] coordinates {(16,0.3435)};
    %\addplot [color=gray, fill] coordinates {(16,0.3431)};
    \addplot [color=violet, fill] coordinates {(16,0.33693)};
    \addplot [color=blue, fill] coordinates {(16,0.2984)};
    %\addplot [color=violet, fill] 
    %table [x=UEs,y=Proposed_conf] {Content/Figures/res_fairness.dat};
\end{axis}
\end{tikzpicture}
    \setlength\abovecaptionskip{0\baselineskip}
    \caption{}\label{fig_se}
\end{subfigure}
\hfill
\begin{subfigure}[t]{0.3\textwidth}
    \centering
    \begin{tikzpicture}[scale = 0.55]
\begin{axis}[
    xlabel = {UEs number $K$},
    ylabel = {Jain's fairness index},
    xtick distance=4,
    ytick distance=0.005,
    xmin=14,xmax=18,
    ymin=0.95,ymax=0.985,
    grid=major,
    yticklabel style={
        /pgf/number format/precision=3,
        /pgf/number format/fixed,
        /pgf/number format/fixed zerofill=true
    },
    label style={font=\large},
    ybar=1pt,
    area legend,
    bar width=12pt,
    legend entries={EX, Proposed US-SA, Proposed Sim. US-SA, MU-PrF, DRL based, mCore+,  RR},
    legend style={
    fill=white,
    fill opacity=0.9,
    text opacity=1,
    legend cell align=left,
    align=left,
    font=\normalsize, % 
    draw=white!15!black,
    at={(0.61,0)}, % 
    anchor=south east,
    legend columns=1}
] 
{EX, Proposed US-SA, Sim. US-SA, MU-PrF, DRL based, mCore+,  RR},
    \addplot [color=black, fill] coordinates {(16,0.982251)};
    \addplot [color=red, fill] coordinates {(16,0.982067)};
    \addplot [color=cyan, fill] coordinates {(16,0.980649)};
    \addplot [color=green, fill] coordinates {(16,0.98)};
    \addplot [color=orange, fill] coordinates {(16,0.9781)};
    \addplot [color=violet, fill] coordinates {(16,0.97872)};
    \addplot [color=blue, fill] coordinates {(16,0.9783)};
\end{axis}
\end{tikzpicture}
    \setlength\abovecaptionskip{0\baselineskip}
    \caption{}\label{fig_fair}
\end{subfigure}
\hfill
\setlength\abovecaptionskip{0\baselineskip}
\caption{(a) Objective value in $(P1)$, (b) normalized system spectral efficiency, and (c) Jain's fairness index when $K=16$.}
\label{fig_satisfaction}
% Channel based on 3GPP TR38.901, where scenario/Scene is Indoor-Mixed office (Open office can be used later)
\end{figure*}
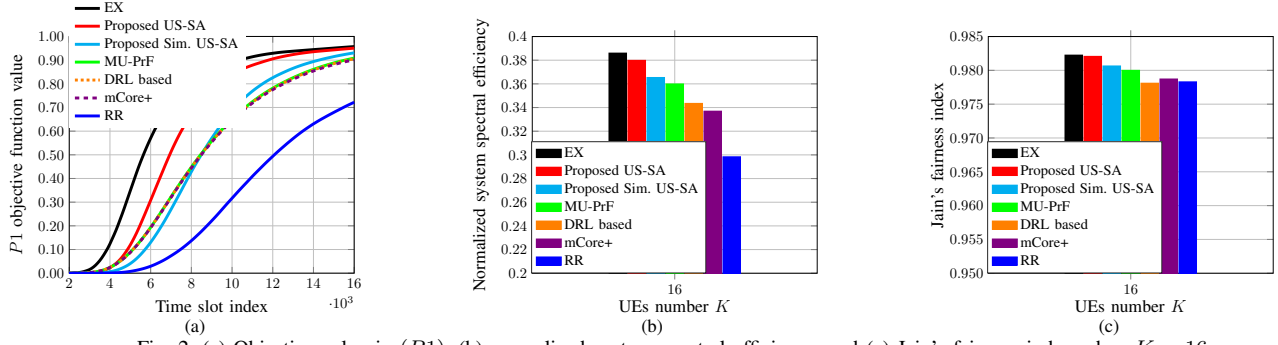

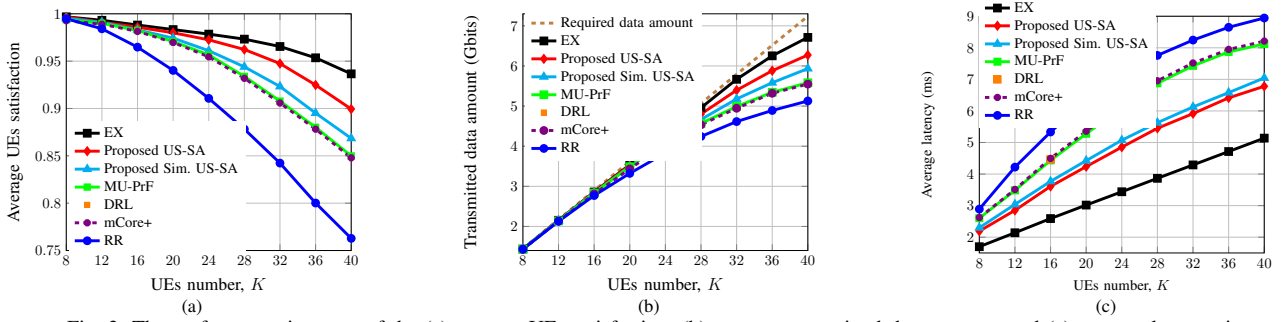
\begin{figure*}[!t]
\centering
\begin{subfigure}[t]{0.31\textwidth}
    \centering
    \begin{tikzpicture}[scale = 0.55]
\begin{axis}[
    xlabel = {UEs number, $K$},
    ylabel = {Average UEs satisfaction},
    xtick distance=4,
    ytick distance=0.05,
    xmin=8,xmax=40,
    ymin=0.75,ymax=1,
    grid=major,
    label style={font=\large},
    legend entries={EX, Proposed US-SA, Proposed Sim. US-SA, MU-PrF, DRL, mCore+, RR},
    legend style={
    draw=none,
    fill=white,
    fill opacity=0,
    text opacity=1,
    legend cell align=left,
    align=left,
    font=\normalsize, % 
    %draw=white!15!black,
    at={(0.63,-0.01)}, % 
    anchor=south east,
    legend columns=1}
]
    \addplot [color=black, line width=2pt, mark=square*, mark options={solid, black}] 
    table [x=UEs,y=EX] {Content/Figures/res_avg_sat.dat};
    %\addplot [color=violet, line width=2pt, mark=*, mark options={solid, violet}] 
    %table [x=UEs,y=Proposed_conf] {Content/Figures/res_avg_sat.dat};
    \addplot [color=red, line width=2pt, mark=diamond*, mark options={solid, red}] 
    table [x=UEs,y=US-SA] {Content/Figures/res_avg_sat.dat};
    \addplot [color=cyan, line width=2pt, mark=triangle*, mark options={solid, cyan}] 
    table [x=UEs,y=Simplified_US-SA] {Content/Figures/res_avg_sat.dat};
    \addplot [color=green, line width=2pt, mark=square*, mark size=1.5pt, mark options={solid, green}] 
    table [x=UEs,y=MU-SPF] {Content/Figures/res_avg_sat.dat};
    \addplot [color=white, mark=square*, mark size=1.75pt, mark options={solid, orange}] coordinates {(16,0.982007)};
    \addplot [color=violet, dashed, line width=2pt, mark=*, mark size=1.5pt, mark options={solid, violet}] 
    table [x=UEs,y=mCore] {Content/Figures/res_avg_sat.dat};
    \addplot [color=blue, line width=2pt, mark=*, mark options={solid, blue}] 
    table [x=UEs,y=RR] {Content/Figures/res_avg_sat.dat};
\end{axis}
\end{tikzpicture}
    \setlength\abovecaptionskip{0\baselineskip}
    \caption{}\label{fig_avg}
\end{subfigure}
\hfill
\begin{subfigure}[t]{0.31\textwidth}
    \centering
    \begin{tikzpicture}[scale = 0.55]
\begin{axis}[
    xlabel = {UEs number, $K$},
    ylabel = {Transmitted data amount (Gbits)},
    xtick distance=4,
    ytick distance=1,
    xmin=8,xmax=40,
    ymin=1.4,ymax=7.3,
    grid=major,
    label style={font=\large},
    legend entries={Required data amount, EX, Proposed US-SA, Proposed Sim. US-SA, MU-PrF, DRL, mCore+, RR},
    legend style={
    draw=none,
    fill=white,
    fill opacity=0,
    text opacity=1,
    legend cell align=left,
    align=left,
    font= \normalsize, % 
    %draw=white!15!black,
    at={(0.63,0.38)}, % 
    anchor=south east,
    legend columns=1}
]
    %\addplot [color=violet, line width=2pt, mark=*, mark options={solid, violet}] 
    %table [x=UEs,y=Proposed_conf] {Content/Figures/res_data_rates.dat};
    \addplot [color=brown, dashed, line width=2pt] 
    coordinates {(8,1.448) (12,2.172) (16,2.896) (20,3.620) (24,4.344) (28,5.068) (32,5.792) (36,6.516) (40,7.240)};
    \addplot [color=black, line width=2pt, mark=square*, mark options={solid, black}] 
    table [x=UEs,y=EX] {Content/Figures/res_data_rates.dat};
    \addplot [color=red, line width=2pt, mark=diamond*, mark options={solid, red}] 
    table [x=UEs,y=US-SA] {Content/Figures/res_data_rates.dat};
    \addplot [color=cyan, line width=2pt, mark=triangle*, mark options={solid, cyan}] 
    table [x=UEs,y=Simplified_US-SA] {Content/Figures/res_data_rates.dat};
    \addplot [color=green, line width=2pt, mark=square*, mark options={solid, green}] 
    table [x=UEs,y=MU-SPF] {Content/Figures/res_data_rates.dat};
    \addplot [color=white, mark=square*, mark size=1.75pt, mark options={solid, orange}] coordinates {(16,2.820309)};
    \addplot [color=violet, dashed, line width=2pt, mark=*, mark options={solid, violet}] 
    table [x=UEs,y=mCore] {Content/Figures/res_data_rates.dat};
    \addplot [color=blue, line width=2pt, mark=*, mark options={solid, blue}] 
    table [x=UEs,y=RR] {Content/Figures/res_data_rates.dat};
\end{axis}
\end{tikzpicture}
    \setlength\abovecaptionskip{0\baselineskip}
    \caption{}\label{fig_rate}
\end{subfigure}
\hfill
\begin{subfigure}[t]{0.31\textwidth}
    \centering
    \begin{tikzpicture}[scale = 0.55]
\begin{axis}[
    xlabel = {UEs number, $K$},
    ylabel = {Average latency (ms)},
    xtick distance=4,
    ytick distance=1,
    xmin=8,xmax=40,
    ymin=1.5,ymax=9,
    grid=major,
    legend entries={EX, Proposed US-SA, Proposed Sim. US-SA, MU-PrF, DRL, mCore+, RR},
    legend style={
    draw=none,
    fill=white,
    fill opacity=0,
    text opacity=1,
    legend cell align=left,
    align=left,
    font= \normalsize, % 
    %draw=white!15!black,
    at={(0.62,0.53)}, % 
    anchor=south east,
    legend columns=1}
]
    \addplot [color=black, line width=2pt, mark=square*, mark options={solid, black}] 
    table [x=UEs,y=EX] {Content/Figures/res_latency.dat};
    \addplot [color=red, line width=2pt, mark=diamond*, mark options={solid, red}] 
    table [x=UEs,y=US-SA] {Content/Figures/res_latency.dat};
    \addplot [color=cyan, line width=2pt, mark=triangle*, mark options={solid, cyan}] 
    table [x=UEs,y=Simplified_US-SA] {Content/Figures/res_latency.dat};
    \addplot [color=green, line width=2pt, mark=square*, mark size=1.75pt, mark options={solid, green}] 
    table [x=UEs,y=MU-SPF] {Content/Figures/res_latency.dat};
    \addplot [color=white, mark=square*, mark size=2.5pt, mark options={solid, orange}] coordinates {(16,4.464803)};
    \addplot [color=violet, dashed, line width=2pt, mark=*, mark size=1.5pt, mark options={solid, violet}] 
    table [x=UEs,y=mCore] {Content/Figures/res_latency.dat};
    \addplot [color=blue, line width=2pt, mark=*, mark options={solid, blue}] 
    table [x=UEs,y=RR] {Content/Figures/res_latency.dat};
    %\addplot [color=violet, line width=2pt, mark=*, mark options={solid, violet}] 
    %table [x=UEs,y=Proposed_conf] {Content/Figures/res_latency.dat};   
\end{axis}
\end{tikzpicture}
    \setlength\abovecaptionskip{0\baselineskip}
    \caption{}\label{fig_lat}
\end{subfigure}
\hfill
\setlength\abovecaptionskip{0\baselineskip}
\caption{The performance in terms of the (a) average UEs satisfaction, (b) average transmitted data amount, and (c) average latency time.}
\label{fig_performance}
\end{figure*}

Figs. \ref{fig_OF}, \ref{fig_se}, and \ref{fig_fair} present the objective in \eqref{P1_eq}, NSE and JFI, respectively, for $K=16$ case. Figs. \ref{fig_avg}, \ref{fig_rate}, and \ref{fig_lat} show the proposed US-SA out-performance with increasing UEs number. MU-PrF, DRL based and mCore+ schemes achieve good performance, but they cannot approach our methods' performance as they aim to optimize different objective, i.e., maximizing sum rate and maintaining UEs fairness, while ours balance maximizing UEs satisfaction and fairness among UEs. US-SA and Sim. US-SA lead versus all competing schemes and approach the EX optimal performance while requiring much less complexity. For example, US-SA requires 16 times lower complexity than the EX approach in the case of 8 associated UEs. Moreover, under overloaded networks with more UEs, e.g., $K=40$, US-SA performance remains stable and better than RR, mCore+ and MU-PrF. Indeed, US-SA is the closest to the optimal EX method, because the size of the sub-grouping matrix $L$ increases linearly with $K$. This proves the scalability of the proposed scheme against increasing associated UEs number.
\begin{table}[!t]
\setlength\tabcolsep{2pt}
\captionsetup{width=0.9\columnwidth}   
\setlength\belowcaptionskip{0\baselineskip}
\caption[]{Simulation parameters.}
\small
\label{sim_parameters}
\begin{tabular}{|p{0.6\columnwidth} | m{0.35\columnwidth}|}
\hline
\textbf{Parameter} & \textbf{Value} \\ [0.3ex] 
\hline
%Carrier frequency & 60 GHz \\
MmWave channel bandwidth, $B$ & 1.825 GHz \\
Bandwidth efficiency, $\eta_B$ & 0.6 \\
Transmit antennas at the BS, $M$ & 16 \\
Multiuser number, $K_u$ & 4 \\
BS transmitted power & 10 dBm\\
Noise power spectral density & -174 dBm/Hz\\
Area dimensions $\times$ height & 25m $\times$ 25m $\times$ 5m\\
UEs heights & 1m\\
Window size, $t_w$ & 1s \\
Numbers of frames, $N_{f}$ & 100 \\
Time slots per frame, $N_{s}$ & 160 \\
One time slot length, $t_{slot}$ & 62.5 $\mu$s \\
%Users required rates, $R^{req}$, in Mbps & 120, 150, 200, 300\\
\hline
\end{tabular}
\end{table}
\begin{comment}
    \begin{subfigure}[t]{0.3\textwidth}
    \centering
    \input{Content/Figures/avg_UEs_full_sat}
    \setlength\abovecaptionskip{0\baselineskip}
    \caption{}\label{fig_full}
\end{subfigure}
\end{comment}
\begin{comment}
    \begin{figure}
    \centering
    \includegraphics[width=1\linewidth]{Content/Figures/fig_OF_4UEs.png}
    \caption{P1 objective function value, UEs = 4.}
    \label{OF_4UEs}
\end{figure}
\begin{figure}
    \centering
    \includegraphics[width=1\linewidth]{Content/Figures/fig_OF_4UEs_zoom.png}
    \caption{System model of 5G network and its applications.}
    \label{OF_4UEs}
\end{figure}
\begin{figure}
    \centering
    \input{Content/Figures/avg_UEs_sats}
    \caption{Average UEs satisfaction of the proposed approach compared to other schemes versus UEs numbers.}
    \label{}
\end{figure}
\end{comment}

\section{Conclusion}
\label{Conclusion}
This paper developed a novel satisfaction-based multiuser scheduling scheme for MU-MIMO networks. Due to the high complexity of the optimal EX solution, US-SA was proposed. US-SA is efficient due to the lower- dimensional sub-grouping matrix employed at each frame, which allows the system to search only through a reduced set to find the best group to be allocated in each slot, and by eliminating satisfied UEs from the scheduling. Our US-SA achieves close to optimal performance and is superior to competing methods. Future extensions will depend on semantic-aware communication metrics alongside UEs demands \cite{SemanticComm_Reply}.

\label{References}
\bibliographystyle{IEEEtran} 
\bibliography{IEEEabrv , Sch_refs} % Entries are in the refs.bib file

\end{document}